\documentclass[prc,preprint,
  showpacs,showkeys,lengthcheck,
  nofootinbib,tightenlines,onecolumn,notitlepage,
  preprintnumbers,superscriptaddress]{revtex4-1}
\usepackage[utf8]{inputenc}
\usepackage{newtxtext}
\usepackage[mathcal]{euscript}

\usepackage[export]{adjustbox}
\usepackage{float}
\usepackage{lipsum}
\usepackage{fixltx2e}

\usepackage{graphicx}
\usepackage[caption=false]{subfig}
\usepackage[usenames,dvipsnames]{xcolor}
\graphicspath{{figures/}}

\usepackage{hyperref}
\hypersetup{colorlinks=true,citecolor=blue,linkcolor=blue,urlcolor=blue}

\usepackage{diagbox}
\usepackage{booktabs}

\usepackage{amsmath,amssymb,amsfonts}
\usepackage{slashed}
\usepackage{bm}

\begin{document}

\title{
  Forces within hadrons on the light front
}

\author{Adam Freese}
\email{afreese@uw.edu}
\address{Department of Physics, University of Washington, Seattle, WA 98195, USA}

\author{Gerald A. Miller}
\email{miller@uw.edu}
\address{Department of Physics, University of Washington, Seattle, WA 98195, USA}

\begin{abstract}
  In this work, we find the light front densities for momentum and forces,
  including pressure and shear forces, within hadrons.
  This is achieved by deriving relativistically correct expressions
  relating these densities to the gravitational form factors
  $A(t)$ and $D(t)$ associated with the energy momentum tensor.
  The derivation begins from the fundamental definition of density in a
  quantum field theory, namely the expectation value of a local operator
  within a spatially-localized state.
  We find that it is necessary to use the light front formalism
  to define a density that corresponds to internal hadron structure.
  When using the instant form formalism,
  it is impossible to remove the spatial extent of the hadron wave function from
  any density, and---even within instant form dynamics---one
  does not obtain a Breit frame Fourier transform
  for a properly defined density.
  Within the front formalism, we derive new expressions for various mechanical
  properties of hadrons, including the mechanical radius,
  as well as for stability conditions.
  The multipole ansatz for the form factors is used as an example
  to illustrate all of these findings.
\end{abstract}

\preprint{NT@UW-21-02}

\maketitle


\section{Introduction}

In the last few years, the energy momentum tensor (EMT) has become a major
focus of both theoretical and experimental efforts in hadron physics.
The proton mass puzzle~\cite{Ji:1994av,Ji:1995sv,Lorce:2017xzd,Hatta:2018sqd,Rodini:2020pis,Metz:2020vxd}
and proton spin puzzle~\cite{Ashman:1987hv,Ji:1996ek,Leader:2013jra}
are major motivators for this focus.
Matrix elements of the EMT between plane wave states
define gravitational form factors~\cite{Kobzarev:1962wt},
which provide information about both the quark-gluon decomposition
and the spatial distribution of energy, momentum, and angular momentum.
These form factors are in principle accessible through high-energy reactions
such as deeply virtual Compton
scattering~\cite{Ji:1996nm,Radyushkin:1997ki,Kriesten:2019jep}
at facilities such as Jefferson
Lab~\cite{Defurne:2015kxq,Jo:2015ema,Hattawy:2019rue}
and the upcoming Electron Ion Collider~\cite{Accardi:2012qut},
as well as in $\gamma\gamma^*\rightarrow\mathrm{hadrons}$
at Belle~\cite{Kumano:2017lhr}.

Investigations into both the EMT and into deeply virtual Compton scattering
have led to the discovery~\cite{Polyakov:1999gs}
of an additional form factor $D(t)$,
called the ``D-term'' or ``Druck-term''~\cite{Polyakov:2018rew},
which does not encode a conserved current,
but instead contains information about the
spatial distribution of
forces within the hadron~\cite{Polyakov:2002yz,Polyakov:2018zvc}.
The D-term has attracted a lot of interest
lately~\cite{Polyakov:2018zvc,Shanahan:2018nnv,Burkert:2018bqq,Kumericki:2019ddg,Freese:2019bhb,Anikin:2019ufr,Neubelt:2019sou,Varma:2020crx},
and remains largely experimentally unexplored.

Since EMT form factors such as the D-term encode spatial densities within
hadrons, it is important that the relationship between the form factors
and spatial densities is properly derived and understood.
The connection between form factors and spatial densities
has been presented in terms of Fourier
transforms~\cite{Sachs:1962zzc,Miller:2009sg,Polyakov:2018zvc,Lorce:2018egm}.
Improvements~\cite{Burkardt:2002hr,Miller:2007uy,Miller:2009sg,Miller:2018ybm,Lorce:2018egm}
in the literature have been made since the original work~\cite{Sachs:1962zzc}
that used three-dimensional form factors.
Specifically,
three-dimensional Fourier transforms at fixed (instant form) time
have been shown to be incorrect~\cite{Miller:2018ybm}
because of the impossibility of localizing
the center of momentum in three dimensions~\cite{Jaffe:2020ebz}.
Two-dimensional Fourier transforms at fixed light front time
have been used as a correct
alternative~\cite{Burkardt:2002hr,Miller:2007uy,Miller:2009sg}.

The improvements have been applied only to electromagnetic form factors
and the associated charge densities,
but the same concerns are also present for the EMT form factors
and energy, spin, and force densities.
It is thus prudent to consider carefully how to properly associate
spatial densities in hadrons with the EMT.
In this paper,
we derive---rather than postulate---a connection between the EMT form factors
on the one hand, and spatial densities of momentum and forces on the other.
In particular, we derive the two-dimensional light front Fourier transform
as providing this connection.

This work is organized as follows.
In Sec.~\ref{sec:fourier},
we derive the connection between spatial densities associated with an
arbitrary local operator and its matrix elements between plane wave states.
This is done both at fixed (instant form) time
and at fixed light front time.
We find only the latter can define a density that depends only on internal
hadron structure, rather than on irremovable wave function spread.
In Sec.~\ref{sec:classical},
we analyze the classical continuum EMT on the light front in order to obtain
associations between the EMT form factors and specific properties of hadrons
(specifically stresses).
In Sec.~\ref{sec:spatial},
we combine the results of the previous sections to obtain results for the
light front momentum density and the light front stress tensor.
Then,
in Sec.~\ref{sec:models}, we illustrate our results with a few model
examples,
and we conclude and provide an outlook in Sec.~\ref{sec:conclusion}.


\section{Spatial densities and Fourier transforms}
\label{sec:fourier}

Spatial densities in quantum field theories are given by expectation values
of local operators with physically realizable states.
Given a local operator $\hat{\mathcal{O}}(x)$---such as the electromagnetic
current or the energy-momentum tensor---and a physically-realizable
one-hadron state $|\Psi\rangle$,
the relevant density is given by
$\langle\Psi|\hat{\mathcal{O}}(x)|\Psi\rangle$.

Form factors are also associated with the local operator $\hat{\mathcal{O}}(x)$
through the matrix element between distinct one-particle plane wave states,
namely through
$\langle p',\lambda'|\hat{\mathcal{O}}(0)|p,\lambda\rangle$.
It is common in the hadron physics literature
to associate the plane wave matrix element with spatial densities using a
Fourier transform, with spacelike components of $\Delta=p'-p$ as the conjugate
momentum variable~\cite{Sachs:1962zzc,Miller:2009sg,Polyakov:2018zvc,Lorce:2018egm}.
In particular, two varieties of Fourier transform exist in the literature:
a three-dimensional ``Breit frame'' Fourier transform
at fixed instant form time $x^0=0$
(see e.g., Refs.~\cite{Sachs:1962zzc,Polyakov:2018zvc,Lorce:2018egm}),
and a two-dimensional Fourier transform over transverse coordinates
at fixed light front time $x^+=0$
(see e.g., Refs.~\cite{Burkardt:2002hr,Miller:2007uy,Miller:2009sg}).
The validity of the Breit frame transform as a true spatial density
has been called into
question~\cite{Miller:2018ybm,Jaffe:2020ebz}.

In this section, we shall derive the correct association between the
matrix elements
$\langle p',\lambda'|\hat{\mathcal{O}}(0)|p,\lambda\rangle$
and the actual field-theoretic spatial densities
$\langle\Psi|\hat{\mathcal{O}}(x)|\Psi\rangle$
for one-hadron states,
using both fixed light front time and fixed instant form time.
We will discuss under what conditions the former can be simplified to
an ordinary two-dimensional Fourier transform over transverse momentum transfer,
and will demonstrate that the latter does not become a
``Breit frame'' Fourier transform.

\subsection{Using the front form}

We shall begin with the light front case.
The light front coordinates are defined so that
$x^\pm = \frac{1}{\sqrt{2}}\Big(x^0 \pm x^3\Big)$.
The density is defined at fixed light front time $x^+=0$,
and the null spatial coordinate $x^-$ will be integrated out,
leaving a two-dimensional spatial density over transverse coordinates
$\mathbf{x}_\perp$.
To start, the following completeness relation for one-hadron states:
\begin{align}
  \sum_\lambda
  \int \frac{\mathrm{d}p^+\mathrm{d}^2\mathbf{p}_\perp}{2p^+(2\pi)^3}
  |p^+,\mathbf{p},\lambda\rangle\langle p^+,\mathbf{p}_\perp,\lambda|
  =
  1
\end{align}
is inserted twice into
$\langle\Psi|\hat{\mathcal{O}}(x)|\Psi\rangle$,
giving:
\begin{align}
  \langle\Psi|\hat{\mathcal{O}}(x)|\Psi\rangle
  =
  \sum_{\lambda\lambda'}
  \int \frac{\mathrm{d}p^+\mathrm{d}^2\mathbf{p}_\perp}{2p^+(2\pi)^3}
  \int \frac{\mathrm{d}p'^+\mathrm{d}^2\mathbf{p}'_\perp}{2p'^+(2\pi)^3}
  \langle\Psi| p'^+,\mathbf{p}'_\perp,\lambda'\rangle
  \langle p'^+,\mathbf{p}'_\perp,\lambda' | \hat{\mathcal{O}}(x) | p^+,\mathbf{p}_\perp,\lambda \rangle
  \langle p^+,\mathbf{p}_\perp,\lambda |\Psi\rangle
  \,.
\end{align}
The spacetime dependence of the local operator $\hat{\mathcal{O}}(x)$
is given by
$\hat{\mathcal{O}}(x) = e^{i\hat{P}\cdot x}\hat{\mathcal{O}}(0)e^{-i\hat{P}\cdot x}$,
and the plane wave states surrounding $\hat{\mathcal{O}}(x)$ are eigenstates of $\hat{P}$.
Using light front coordinates at fixed $x^+=0$ gives
$\hat{P}\cdot x = \hat{P}^+x^- - \hat{\mathbf{P}}_\perp\cdot\mathbf{x}_\perp$.
If we define a change of variables
$P = \frac{1}{2}\big(p+p'\big)$ and $\Delta = p'-p$, then we find that:
\begin{multline}
  \langle\Psi|\hat{\mathcal{O}}(x^+=0,x^-,\mathbf{x}_\perp)|\Psi\rangle
  =
  \sum_{\lambda\lambda'}
  \int \frac{\mathrm{d}P^+\mathrm{d}^2\mathbf{P}_\perp}{2(2\pi)^3}
  \int \frac{\mathrm{d}\Delta^+\mathrm{d}^2\boldsymbol{\Delta}_\perp}{2(2\pi)^3}
  \frac{1}{P^++\frac{\Delta^+}{2}}
  \frac{1}{P^+-\frac{\Delta^+}{2}}
  \\
  \langle\Psi| p'^+,\mathbf{p}'_\perp,\lambda'\rangle
  \langle p'^+,\mathbf{p}'_\perp,\lambda' | \hat{\mathcal{O}}(0) | p^+,\mathbf{p}_\perp,\lambda \rangle
  \langle p^+,\mathbf{p}_\perp,\lambda |\Psi\rangle
  e^{i(\Delta^+x^- - \boldsymbol{\Delta}_\perp\cdot\mathbf{x}_\perp)}
  \,.
\end{multline}
In order to remove the dependence of the density on the state that the hadron
is prepared in,
it must be possible to factorize the $P$ and $\Delta$ dependence of the integrand.
The factors
$\left(P^++\frac{\Delta^+}{2}\right)^{-1}$
and
$\left(P^+-\frac{\Delta^+}{2}\right)^{-1}$
impede this factorization unless $\Delta^+=0$,
which can be achieved by integrating out $x^-$.
Doing so gives us:
\begin{multline}
  \label{eqn:232}
  \int \mathrm{d}x^- \,
  \langle\Psi|\hat{\mathcal{O}}(x^+=0,x^-,\mathbf{x}_\perp)|\Psi\rangle
  =
  \sum_{\lambda\lambda'}
  \int \frac{\mathrm{d}P^+\mathrm{d}^2\mathbf{P}_\perp}{2P^+(2\pi)^3}
  \int \frac{\mathrm{d}^2\boldsymbol{\Delta}_\perp}{(2\pi)^2}
  \\
  \langle\Psi| P^+,\mathbf{p}'_\perp,\lambda'\rangle
  \frac{
    \langle P^+,\mathbf{p}'_\perp,\lambda' |
    \hat{\mathcal{O}}(0)
    | P^+,\mathbf{p}_\perp,\lambda \rangle
  }{2P^+}
  \langle P^+,\mathbf{p}_\perp,\lambda |\Psi\rangle
  e^{-i\boldsymbol{\Delta}_\perp\cdot\mathbf{x}_\perp}
  \,.
\end{multline}
This is the general expression for an arbitrary hadron state $|\Psi\rangle$.
For such a general state, the density associated with
$\hat{\mathcal{O}}(x)$ depends not only on the hadron's internal structure,
but also on its wave function spread.

To obtain a density depending only on internal structure,
a wave function that is completely localized at the origin should be used.
This must be done carefully,
since absolute position states do not exist in
Hilbert space~\cite{vonNeumann:1955book}.
In particular, localization is achieved by considering wave packets with
arbitrarily narrow width in the position representation,
though this width must be kept finite until after all
other calculations have been performed~\cite{Miller:2018ybm}.
A specific example---which is also a $P^+$ plane wave state,
and a state with definite light front helicity $\Lambda$---is given by:
\begin{align}
  \label{eqn:wf:lf}
  \langle p^+,\mathbf{p}_\perp,\lambda |\Psi\rangle
  =
  \sqrt{2\pi} (2\sigma) e^{-\sigma^2 \mathbf{p}_\perp^2}
  \,
  \sqrt{2p^+(2\pi) \delta_\epsilon(p^+-P^+) }
  \,
  \delta_{\lambda\Lambda}
  \,,
\end{align}
where $\delta_\epsilon(x)$ is an arbitrary representation of the
Dirac delta function
(such that $\delta_\epsilon(x)\rightarrow\delta(x)$
as $\epsilon\rightarrow0$).
Ultimately, the $\sigma\rightarrow0$ limit should not be taken until
after all momentum integrals have been performed,
except in special cases such as when the dominated convergence theorem
allows the limit to be commuted with integration.

Using the wave function of Eq.~(\ref{eqn:wf:lf}) in Eq.~(\ref{eqn:232}) gives:
\begin{multline}
  \int \mathrm{d}x^- \,
  \langle\Psi|\hat{\mathcal{O}}(x^+=0,x^-,\mathbf{x}_\perp)|\Psi\rangle
  =
  (2\pi)(2\sigma)^2
  \int \frac{\mathrm{d}^2\mathbf{P}_\perp}{(2\pi)^2}
  e^{-2\sigma^2 \mathbf{P}_\perp^2}
  \\
  \int \frac{\mathrm{d}^2\boldsymbol{\Delta}_\perp}{(2\pi)^2}
  \frac{
    \langle P^+,\mathbf{p}'_\perp,\Lambda |
    \hat{\mathcal{O}}(0)
    | P^+,\mathbf{p}_\perp,\Lambda \rangle
  }{2P^+}
  e^{-i\boldsymbol{\Delta}_\perp\cdot\mathbf{x}_\perp}
  e^{-\frac{\sigma^2}{2} \boldsymbol{\Delta}_\perp^2}
  \,.
\end{multline}
This is the most general possible result for the $x^-$-integrated
density at $x^+=0$.
Whether it is possible to proceed from here depends on the form
of the specific matrix element
$
    \langle P^+,\mathbf{p}'_\perp,\lambda' |
    \hat{\mathcal{O}}(0)
    | P^+,\mathbf{p}_\perp,\lambda \rangle
$.
In particular, the possibility of proceeding depends on whether
the Lorentz structures in this matrix elements contain any factors
of $\mathbf{P}_\perp$.
If there are no such factors, it is possible to do the
$\mathbf{P}_\perp$ integration, giving:
\begin{align}
  \label{eqn:fourier:lf}
  \int \mathrm{d}x^- \,
  \langle\Psi|\hat{\mathcal{O}}(x^+=0,x^-,\mathbf{x}_\perp)|\Psi\rangle
  =
  \int \frac{\mathrm{d}^2\boldsymbol{\Delta}_\perp}{(2\pi)^2}
  \frac{
    \langle P^+,\mathbf{p}'_\perp,\Lambda |
    \hat{\mathcal{O}}(0)
    | P^+,\mathbf{p}_\perp,\Lambda \rangle
  }{2P^+}
  e^{-i\boldsymbol{\Delta}_\perp\cdot\mathbf{x}_\perp}
  e^{-\frac{\sigma^2}{2} \boldsymbol{\Delta}_\perp^2}
  \,,
\end{align}
which is the first major result of this section.

On the other hand, if there are additional factors of
$\mathbf{P}_\perp$, the resulting integral will contain a factor
$\sigma^{-1}$ or worse, causing the result to diverge in the
$\sigma\rightarrow0$ limit.
This essentially occurs because of complementarity:
an observable that depends on $\mathbf{P}_\perp$ cannot be well-defined
in the limit of the hadron's transverse position being arbitrarily well-known.

It is of course possible to calculate finite spatial densities for
any current, provided we avoid the limit of arbitrary spatial localization.
However, if we wish for our densities to encode the internal structure
of hadrons rather than wave function spread,
we are constrained to calculating spatial densities for currents
that do not depend on $\mathbf{P}_\perp$.
As we shall see in Sec.~\ref{sec:classical},
the spatial components of the energy momentum tensor are not suitable
candidates for defining such an internal spatial density.
However, we will also find that matrix elements of the pure stress tensor---that
is, the local values of the stress tensor in a comoving frame, which
encodes pressure and shear---do not depend on $\mathbf{P}_\perp$, and therefore
the stress tensor is a suitable candidate for internal spatial densities.

\subsection{Using the instant form}

Let us next briefly
consider spatial densities at a fixed instant form time $x^0=0$.
This time we use the completeness relation:
\begin{align}
  \sum_\lambda
  \int \frac{\mathrm{d}^3\mathbf{p}}{2E_p(2\pi)^3}
  |\mathbf{p},\lambda\rangle\langle \mathbf{p},\lambda|
  =
  1
  \,,
\end{align}
inserted twice into the expectation value
$\langle\Psi|\hat{\mathcal{O}}(x)|\Psi\rangle$
to find:
\begin{align}
  \langle\Psi|\hat{\mathcal{O}}(x)|\Psi\rangle
  =
  \sum_{\lambda\lambda'}
  \int \frac{\mathrm{d}^3\mathbf{p}}{2E_p(2\pi)^3}
  \int \frac{\mathrm{d}^3\mathbf{p}'}{2E_p'(2\pi)^3}
  \langle\Psi| \mathbf{p}',\lambda'\rangle
  \langle \mathbf{p}',\lambda' | \hat{\mathcal{O}}(x) | \mathbf{p},\lambda \rangle
  \langle \mathbf{p},\lambda |\Psi\rangle
  \,.
\end{align}
Using the change of variables
$\mathbf{P} = \frac{1}{2}\big(\mathbf{p}+\mathbf{p}'\big)$
and
$\boldsymbol{\Delta} = \mathbf{p}'-\mathbf{p}$,
and the spatial dependence
$\hat{\mathcal{O}}(x) = e^{i\hat{P}\cdot x}\hat{\mathcal{O}}(0)e^{-i\hat{P}\cdot x}$,
at fixed time $x^0=0$, we find:
\begin{align}
  \langle\Psi|\hat{\mathcal{O}}(\mathbf{x},x^0=0)|\Psi\rangle
  =
  \sum_{\lambda\lambda'}
  \int \frac{\mathrm{d}^3\mathbf{P}}{(2\pi)^3}
  \int \frac{\mathrm{d}^3\boldsymbol{\Delta}}{(2\pi)^3}
  \frac{1}{4E_pE_p'}
  \langle\Psi| \mathbf{p}',\lambda'\rangle
  \langle \mathbf{p}',\lambda' | \hat{\mathcal{O}}(0) | \mathbf{p},\lambda \rangle
  \langle \mathbf{p},\lambda |\Psi\rangle
  e^{-i\boldsymbol{\Delta}\cdot\mathbf{x}}
  \,.
\end{align}
It is controversial whether spatial localization at fixed $x^0$ is possible,
and proposed localization methods typically redefine the concept of
localization to admit some wave function
spread~\cite{Newton:1949cq,Kalnay:1967zz,Pavsic:2018vbs}.
As a loose localization procedure in momentum space,
we can use the following wave function~\cite{Jaffe:2020ebz}:
\begin{align}
  \label{eqn:wf:instant}
  \langle \mathbf{p},\lambda |\Psi\rangle
  =
  \sqrt{2E_p}
  \,
  (2\pi)^{3/4} (2\sigma)^{3/2}
  e^{-\sigma^2\mathbf{p}^2}
  \delta_{\lambda\Lambda}
  \,.
\end{align}
Using this wave function, we find:
\begin{align}
  \label{eqn:fourier:instant}
  \langle\Psi|\hat{\mathcal{O}}(\mathbf{x},x^0=0)|\Psi\rangle
  =
  (2\pi)^{3/2} (2\sigma)^3
  \int \frac{\mathrm{d}^3\mathbf{P}}{(2\pi)^3}
  \int \frac{\mathrm{d}^3\boldsymbol{\Delta}}{(2\pi)^3}
  \sqrt{\frac{1}{4E_pE_p'}}
  \,
  e^{-2\sigma^2\mathbf{P}^2}
  \langle \mathbf{p}',\Lambda | \hat{\mathcal{O}}(0) | \mathbf{p},\Lambda \rangle
  e^{-\frac{\sigma^2}{2}\boldsymbol{\Delta}^2}
  e^{-i\boldsymbol{\Delta}\cdot\mathbf{x}}
  \,.
\end{align}
The energy factors
$E_p = \sqrt{M^2 + \left(\mathbf{P}-\frac{1}{2}\boldsymbol{\Delta}\right)^2}$
and
$E_p' = \sqrt{M^2 + \left(\mathbf{P}+\frac{1}{2}\boldsymbol{\Delta}\right)^2}$
do not factorize in their $\mathbf{P}$ and $\boldsymbol{\Delta}$ dependence.
Factorizing the integrand into a $\mathbf{P}$-dependent piece and a
$\boldsymbol{\Delta}$-dependent piece would require setting
$\boldsymbol{\Delta}=0$, which can be achieved by integrating out $\mathbf{x}$.
However, this would preclude obtaining a density.

As previously found in Refs.~\cite{Miller:2018ybm,Jaffe:2020ebz},
Eq.~(\ref{eqn:fourier:instant}) tells us that dependence on the hadron's
wave function cannot be removed or even factored out of the spatial density
when using fixed instant form time.
Moreover, the expression obtained in Eq.~(\ref{eqn:fourier:instant})
does not resemble the conventional Breit frame Fourier transform at all.
One would need to eliminate the $\mathbf{P}$ integral while setting
$\mathbf{P}=0$ in order to obtain the Breit frame transform,
but such a procedure is incompatible with the spatially-localized state
that was used.
States localized in momentum space---i.e., plane waves---would need to
be used to set $\mathbf{P}=0$, but as found in Ref.~\cite{Miller:2018ybm},
such a state produces singular densities with infinite radii since the state
is completely delocalized.

Therefore, the Breit frame Fourier transform does not give a spatial density.
On the other hand,
Eq.~(\ref{eqn:fourier:instant}) does give a valid spatial density, but
this density does not correspond strictly to internal structure of the hadron.
Wave function spread will always be present in any density defined
at fixed $x^0$.
It is thus preferable to use the light front density of
Eq.~(\ref{eqn:fourier:lf}) to describe hadron structure,
since wave function spread can be factored out---and in some cases,
eliminated entirely.


\section{The light front stress tensor}
\label{sec:classical}

Identifying classical mechanical concepts
such as pressure and shear forces
in inherently quantum mechanical systems such as hadrons is difficult.
In practice, ``pressure'' is identified within hadrons by comparing
matrix elements of the EMT for hadrons states to a continuum EMT.
The matrix element is typically evaluated using plane wave states
in the Breit frame, with $\mathbf{P}=0$,
so that the stress tensor---which is identified with the spatial
components of the EMT---does not depend on the bulk velocity of the hadron.

However, we have established in Sec.~\ref{sec:fourier}
that there is no connection between Breit frame matrix elements
and actual spatial densities.
Moreover, a properly defined spatial density involves an integral over all
values of the transverse momentum $\mathbf{P}_\perp$,
meaning that the stress tensor contains contributions
from the bulk flow of energy and momentum,
rather than just forces internal to the hadron.
It is necessary to isolate the part of the stress tensor corresponding
strictly to internal forces,
which we call the ``pure stress tensor.''
We shall proceed to consider how this can be done.

Since the classical EMT is symmetric,
we use the symmetric,
Belinfante-improved EMT~\cite{Belinfante:1939emt}
on the field theoretic side of the comparison.

\subsection{Continuum EMT on the light front}

Since spatial densities encoding internal structure of hadrons can only
be defined at equal light front time, with $x^-$ integrated out,
we shall begin by considering the general properties
of a classical EMT under these conditions.
The Poincar\'e group has a Galilean
subgroup~\cite{Dirac:1949cp,Susskind:1967rg,Brodsky:1997de}
that leaves the foliation of spacetime into $x^+$ slices invariant.
Besides the Hamiltonian $P^-$, which generates $x^+$ translations,
the remaining generators of the subgroup leave $x^+=\mathrm{fixed}$ invariant.
Moreover, transformations in this Galilean subgroup have the special property
that the $+$ and transverse components of transformed tensors do not depend
on the $-$ components in the original frame.

It is therefore prudent to proceed considering only the $+,1,2$ components
of the EMT and of other tensors.
We thus proceed with the inherently $(2+1)$-dimensional quantity:
\begin{align}
  T^{\mu\nu}_{\mathrm{LF}}(x^+,\mathbf{x}_\perp)
  =
  \int \mathrm{d}x^- \,
  T^{\mu\nu}(x^+,x^-,\mathbf{x}_\perp)
  \,,
  \qquad : \,
  \mu,\nu = +,1,2
  \,.
\end{align}
The EMT is made up of two pieces:
a flow tensor
$V^{\mu\nu}_{\mathrm{LF}}(x^+,\mathbf{x}_\perp)$
that encodes the local motion of the continuum material,
and the pure stress tensor
$S^{\mu\nu}_{\mathrm{LF}}(x^+,\mathbf{x}_\perp)$
that encodes mechanical forces:
\begin{align}
  T^{\mu\nu}_{\mathrm{LF}}(x^+,\mathbf{x}_\perp)
  =
  V^{\mu\nu}_{\mathrm{LF}}(x^+,\mathbf{x}_\perp)
  +
  S^{\mu\nu}_{\mathrm{LF}}(x^+,\mathbf{x}_\perp)
  \,.
\end{align}
The pure stress tensor evaluated at $\mathbf{x}_\perp$
is what the EMT evaluates to in a frame that's comoving
with the material at $\mathbf{x}_\perp$~\cite{Fetter:1980str}.

To better understand the flow and pure stress tensors,
let us first consider a small element of material at transverse rest\footnote{
  By ``transverse rest,'' we mean that once $x^-$ is integrated out,
  the net momentum in the transverse plane is zero.
}.
By Noether's theorem, the EMT components
$T^{+\nu}_{\mathrm{LF}}(x^+,\mathbf{x}_\perp)$
encode the densities of the momentum components $P^\nu$,
which for the material at transverse rest gives
$T^{+i}_{\mathrm{LF},\mathrm{rest}}(x^+,\mathbf{x}_\perp)=0$.
On the other hand, the light front momentum ($P^+$) density is given by:
\begin{align}
  \varepsilon(x^+,\mathbf{x}_\perp)
  =
  T^{++}_{\mathrm{LF},\mathrm{rest}}(x^+,\mathbf{x}_\perp)
  \,.
\end{align}
In this case, the flow tensor is defined by:
\begin{align}
  V^{\mu\nu}_{\mathrm{LF},\mathrm{rest}}(x^+,\mathbf{x}_\perp)
  =
  \left[
    \begin{array}{ccc}
      \varepsilon(x^+,\mathbf{x}_\perp) & 0 & 0 \\
      0 & 0 & 0 \\
      0 & 0 & 0
    \end{array}
    \right]
  \,.
\end{align}
The pure stress tensor
$S^{\mu\nu}_{\mathrm{LF}}(x^+,\mathbf{x}_\perp)$
is defined through the remaining spatial components of 
$T^{\mu\nu}_{\mathrm{LF}}(x^+,\mathbf{x}_\perp)$,
and can be written out in terms of components as:
\begin{align}
  S^{\mu\nu}_{\mathrm{LF},\mathrm{rest}}(x^+,\mathbf{x}_\perp)
  =
  \left[
    \begin{array}{ccc}
      0 & 0 & 0 \\
      0 & S_{11}(x^+,\mathbf{x}_\perp) & S_{12}(x^+,\mathbf{x}_\perp) \\
      0 & S_{21}(x^+,\mathbf{x}_\perp) & S_{22}(x^+,\mathbf{x}_\perp)
    \end{array}
    \right]
  \,.
\end{align}
The pure stress tensor has only spatial components in the transverse rest frame
of the material element,
and under transformations within the light front Galilean subgroup---such
as transverse boosts---this property remains invariant.
Thus $S^{++} = S^{+i} = S^{i+} = 0$ is true in all frames.

On the other hand,
$V^{\mu\nu}_{\mathrm{LF}}(x^+,\mathbf{x}_\perp)$
is not invariant under transverse boosts,
but instead has the generic form:
\begin{align}
  V^{\mu\nu}_{\mathrm{LF}}(x^+,\mathbf{x}_\perp)
  =
  u^\mu
  u^\nu
  \varepsilon(x^+,\mathbf{x}_\perp)
  \,,
\end{align}
where $u^\mu$ is the light front velocity of the material element,
with $u^+=1$ and $P^+\mathbf{u}_\perp = \mathbf{p}_\perp$.

A general continuum material consists of many small elements that
may be in motion relative to each other.
The general form of the continuum EMT on the light front
is thus obtained by making the light front velocity
a function of space and time:
\begin{align}
  \label{eqn:EMT:classical}
  T^{\mu\nu}_{\mathrm{LF}}(x^+,\mathbf{x}_\perp)
  =
  u^\mu(x^+,\mathbf{x}_\perp)
  u^\nu(x^+,\mathbf{x}_\perp)
  \varepsilon(x^+,\mathbf{x}_\perp)
  +
  S^{\mu\nu}_{\mathrm{LF}}(x^+,\mathbf{x}_\perp)
  \,.
\end{align}
It's worth noting that rotational motion of the material can also
be encoded through the velocity field $u^\mu(x^+,\mathbf{x}_\perp)$.
However, since this is a classical description,
intrinsic quark spin---which, in quantum field theories,
is encoded in the antisymmetric contribution to the
EMT~\cite{Leader:2013jra}---cannot
be accommodated by Eq.~(\ref{eqn:EMT:classical}).

We shall now proceed to consider properties of 
$T^{\mu\nu}_{\mathrm{LF}}(x^+,\mathbf{x}_\perp)$,
$V^{\mu\nu}_{\mathrm{LF}}(x^+,\mathbf{x}_\perp)$,
and
$S^{\mu\nu}_{\mathrm{LF}}(x^+,\mathbf{x}_\perp)$.
Firstly, the unintegrated EMT obeys a continuity equation:
\begin{align}
  \partial_\mu T^{\mu\nu}(x) = 0
  \,,
\end{align}
as a consequence of Noether's theorem.
If the EMT vanishes at spatial infinity,
then the $x^-$-integrated EMT also obeys a continuity equation:
\begin{align}
  \label{eqn:conservation:LF}
  \partial_\mu T_{\mathrm{LF}}^{\mu\nu}(x^+,\mathbf{x}_\perp) = 0
  \,.
\end{align}
The pure stress tensor are subject
to Cauchy's first law of mechanics~\cite{Irgens:2008con}:
\begin{align}
  \nabla_i S_{\mathrm{LF}}^{ij}(x^+,\mathbf{x}_\perp)
  =
  -
  \mathbf{F}_\perp^j(x^+,\mathbf{x}_\perp)
  \,,
\end{align}
which can be seen to follow from Eq.~(\ref{eqn:conservation:LF})
with the definition:
\begin{align}
  \mathbf{F}_\perp(x^+,\mathbf{x}_\perp)
  =
  \partial_+ \Big[
    \varepsilon(x^+,\mathbf{x}_\perp)
    \mathbf{u}_\perp(x^+,\mathbf{x}_\perp)
    \Big]
  \,,
\end{align}
which is effectively a statement of Newton's second law:
a net force produces a change in momentum with time.
Thus,
$\nabla_i S_{\mathrm{LF}}^{ij}(x^+,\mathbf{x}_\perp)$
gives the local force acting on elements of the material.

For a system in equilibrium---such as an isolated hadron---the
net force at all locations should be zero\footnote{
  The force on quarks and the force on gluons may each be non-zero,
  but need to sum to zero.
}.
Since $S^{+\nu}=0$, the equilibrium condition can also be written:
\begin{align}
  \partial_\mu S_{\mathrm{LF}}^{\mu\nu}(x^+,\mathbf{x}_\perp)
  =
  0
  \,,
\end{align}
which when combined with the continuity equation for the EMT,
also gives:
\begin{align}
  \partial_\mu V_{\mathrm{LF}}^{\mu\nu}(x^+,\mathbf{x}_\perp)
  =
  0
  \,.
\end{align}
This means that, besides the EMT as a whole,
the flow tensor and pure stress tensor are separately conserved quantities
in an equilibrium system.

\subsection{Gravitational form factors}

When using inherently classical formulas such as Eq.~(\ref{eqn:EMT:classical})
in a quantum mechanical context,
the quantities involved should be understood as expectation values
using physical states.
As found in Sec.~\ref{sec:fourier}, these expectation values can be expressed
as Fourier transforms of matrix elements
$\langle p',\lambda|\hat{\mathcal{O}}(x)|p,\lambda\rangle$
between plane wave states.

The matrix elements
$\langle p',\lambda|T^{\mu\nu}(0)|p,\lambda\rangle$
in particular define gravitational form factors.
For spin-zero hadrons, the standard decomposition
is~\cite{Polyakov:2018zvc}:
\begin{align}
  \label{eqn:emt:spin0}
  \langle p' |
  T^{\mu\nu}(0)
  |p\rangle
  =
  2 P^\mu P^\nu A(t)
  + \frac{\Delta^\mu\Delta^\nu-\Delta^2g^{\mu\nu}}{2} D(t)
  \,,
\end{align}
where $t = \Delta^2$ is the Lorentz-invariant squared momentum transfer;
while for spin-half particles:
\begin{align}
  \langle p',\lambda |
  T^{\mu\nu}(0)
  |p,\lambda\rangle
  =
  \bar{u}(p',\lambda) \left\{
    \frac{P^\mu P^\nu}{M} A(t)
    +
    \frac{\Delta^\mu\Delta^\nu-\Delta^2g^{\mu\nu}}{4M} D(t)
    +
    \frac{iP^{\{\mu}\sigma^{\nu\}\rho}\Delta_\rho}{2M} J(t)
    \right\}
  u(p,\lambda)
  \,,
\end{align}
where the curly brackets $\{\}$ signify symmetrization over the indices
between them (without a factor $1/2$).
Using the Kogut-Soper spinors~\cite{Kogut:1969xa} and $\Delta^+=0$\footnote{
  Kogut-Soper spinors give the same results for matrix elements
  as Brodsky-Lepage spinors~\cite{Lepage:1980fj}.
  Table II of Ref.~\cite{Lepage:1980fj} contains various matrix elements
  of these spinors.
  Note that these matrix elements may be different for spinors with
  canonical polarization,
  but light front helicity spinors such as in
  Refs.~\cite{Kogut:1969xa,Lepage:1980fj}
  are more natural to use in the light front formalism.
}
(as happens to be the case when evaluating spatial densities),
we find through explicit evaluation
that the $+,1,2$ components of spin-half matrix element can be rewritten:
\begin{align}
  \label{eqn:emt:spin12}
  \langle p',\lambda |
  T^{\mu\nu}(x^-)
  |p,\lambda\rangle
  =
    2 P^\mu P^\nu A(t)
    +
    \frac{\Delta^\mu\Delta^\nu-\Delta^2g^{\mu\nu}}{2} D(t)
    -
    i\lambda P^{\{\mu}\epsilon^{\nu\}\rho+}\Delta_\rho
    J(t)
  \,,
\end{align}
where the Levi-Civita symbol should be understood as a
three-dimensional quantity with
$\epsilon^{+12} = 1$,
and which we emphasize is only true at $\Delta^+=0$.

A few pertinent properties of the flow and pure stress tensors allow us
to identify terms in Eqs.~(\ref{eqn:emt:spin0},\ref{eqn:emt:spin12})
with
$V^{\mu\nu}_{\mathrm{LF}}(0)$
and
$S^{\mu\nu}_{\mathrm{LF}}(0)$.
First,
since the hadron is in equilibrium,
the flow and pure stress tensors are separately conserved,
meaning the Lorentz structures associated with each should contract with
$\Delta_\mu$ to zero.
The Lorentz structures accompanying $A(t)$, $D(t)$, and $J(t)$ each
separately satisfy this constraint already.
In addition,
the facts that $S^{+i}=S^{i+}=S^{++}=0$
and that $V^{ij}$ should vanish
for the transverse rest state $\mathbf{P}_\perp=0$
allow us to uniquely identify:
\begin{subequations}
  \label{eqn:spatial}
  \begin{align}
    \label{eqn:spatial:flow}
    \langle p',\lambda |
    V^{\mu\nu}_{\mathrm{LF}}(0)
    |p,\lambda\rangle
    &=
    (2\pi) \delta(\Delta^+)
    \left\{
      2 P^\mu P^\nu A(t)
      -
      i\lambda P^{\{\mu}\epsilon^{\nu\}\rho+}\Delta_\rho
      J(t)
      \right\}
    \,,
    \\
    \label{eqn:spatial:stress}
    \langle p',\lambda |
    S^{\mu\nu}_{\mathrm{LF}}(0)
    |p,\lambda\rangle
    &=
    (2\pi) \delta(\Delta^+)
    \,
    \left(\frac{\Delta^\mu\Delta^\nu-\Delta^2g^{\mu\nu}}{2}\right) D(t)
    \,,
  \end{align}
\end{subequations}
where the $J(t)$ term is absent for spin-zero hadrons.
(Note that
$V^{\mu\nu}_{\mathrm{LF}}(0)$
and
$S^{\mu\nu}_{\mathrm{LF}}(0)$
by definition already include an integral over $x^-$,
which is why the factors of $\delta(\Delta^+)$ are present.)

As explained in the discussion surrounding Eq.~(\ref{eqn:fourier:lf}),
one cannot obtain an intrinsic spatial density
(i.e., a density that is finite for spatially localized states)
from a matrix element that involves factors of $\mathbf{P}_\perp^i$.
The pure stress tensor is an inherently good quantity
for defining spatial densities,
but of the flow tensor,
only $V_{\mathrm{LF}}^{++}=T_{\mathrm{LF}}^{++}$ gives a good spatial density.
This of course means that the spatial components of the EMT---which constitute
the entire stress tensor, as conventionally defined
(see Refs.~\cite{Fetter:1980str,Batchelor:2000int,Irgens:2008con})---do
not constitute a good intrinsic spatial density,
but that the pure stress tensor---which corresponds to how the stress tensor
is seen by a comoving observer at each point in the material,
and which corresponds to the $D(t)$ form factor---does constitute a good
spatial density.


\section{Spatial densities of the EMT}
\label{sec:spatial}

As we have seen in Sec.~\ref{sec:fourier}, it is only possible to define
spatial densities which encode the internal structure of hadrons
within the light front formalism.
Moreover, such densities can only be defined for operators whose matrix
elements between plane wave states do not depend on the average transverse
momentum between the states.
When these conditions are met, the spatial density is given by
Eq.~(\ref{eqn:fourier:lf}).

\subsection{Light front momentum density}

It was found in Sec.~\ref{sec:classical} that $T^{++}(x)$ is the only component
of the EMT that defines a good spatial density.
Using Eq.~(\ref{eqn:fourier:lf}),
along with either of the form factor decompositions in
Eqs.~(\ref{eqn:emt:spin0},\ref{eqn:emt:spin12}) gives:
\begin{align}
  \label{eqn:density:p+}
  \varepsilon(\mathbf{x}_\perp)
  \equiv
  \int \mathrm{d}x^-
  \langle \Psi | T^{++}(x) | \Psi \rangle
  =
  P^+
  \int \frac{\mathrm{d}^2\boldsymbol{\Delta}_\perp}{(2\pi)^2}
  A(t)
  e^{-i\boldsymbol{\Delta}_\perp\cdot\mathbf{x}_\perp}
  e^{-\frac{\sigma^2}{2} \boldsymbol{\Delta}_\perp^2}
  \,,
\end{align}
This is the light front  momentum density 
for the + component.

As a momentum density, the integral of $\varepsilon(\mathbf{x}_\perp)$
over all space should give the actual value of the momentum $P^+$.
Since:
\begin{align}
  \int \mathrm{d}^2\mathbf{x}_\perp \,
  \varepsilon(\mathbf{x}_\perp)
  =
  P^+ A(0)
  \,,
\end{align}
this imposes the well-known momentum sum rule $A(0)=1$.

The $\mathbf{x}_\perp^2$-weighted integral of $\varepsilon(\mathbf{x}_\perp)$,
when the net momentum $P^+$ is divided out,
defines the square of a light front momentum radius:
\begin{align}
  \label{eqn:radius:p+}
  \langle x_\perp^2 \rangle_{\mathrm{mom}}
  \equiv
  \frac{1}{P^+}
  \int \mathrm{d}^2\mathbf{x}_\perp \,
  \mathbf{x}_\perp^2
  \varepsilon(\mathbf{x}_\perp)
  =
  4
  \frac{\mathrm{d}A(t)}{\mathrm{d}t}\bigg|_{t=0}
  \,.
\end{align}
Thus, the light front formalism provides a physical interpretation for the
slope of $A(t)$.
It is worth remarking
that the pion ``mass radius'' extracted from Ref.~\cite{Kumano:2017lhr}
is actually $\sqrt{3/2}$ times the $P^+$ radius.

Although it describes a different quantity---energy rather than
momentum---the most conceptually adjacent concept provided by the Breit frame
formalism is its energy radius~\cite{Polyakov:2018zvc,Freese:2019bhb}:
\begin{align}
  \langle r^2 \rangle_{\mathrm{Breit}}
  =
  6
  \frac{\mathrm{d}A(t)}{\mathrm{d}t}\bigg|_{t=0}
  -
  \frac{3}{4M^2} \Big\{
    A(0) - 2J(0) + 2D(0) \Big\}
  \,,
\end{align}
where $J(0)=0$ for spin-zero hadrons
and $J(0)=\frac{1}{2}$ for spin-half hadrons.
As a theoretical quantity for understanding hadron structure,
the light front $P^+$ radius has several virtues over the
Breit frame energy radius.
The most obvious of these virtues should be that the light front momentum radius
is the radius of an actual spatial density.
However, there are several other peculiarities
exhibited by the Breit frame radius.

For a spin-zero point particle,
one has $A(0)=1$ and $D(0)=-1$~\cite{Hudson:2017oul}.
The Breit frame energy radius does not vanish, but is instead given by
$\langle r^2 \rangle_{\mathrm{Breit}} = \frac{3}{4M^2}$,
while the light front momentum radius vanishes as expected.
Additionally, for massless hadrons with either spin zero or spin half,
the Breit frame radius is infinite,
while the light front momentum radius remains finite.

Another minor benefit of the light front momentum radius is that it is simply
given by (a number times) the slope of the form factor $A(t)$.

\subsection{Light front pure stress tensor}

As seen in Sec.~\ref{sec:classical}, the spatial components of the EMT do not
define a good spatial density,
but the pure stress tensor does define a good density.
In particular,
using Eq.~(\ref{eqn:spatial:stress})
in Eq.~(\ref{eqn:fourier:lf}) gives:
\begin{align}
  \label{eqn:Sij}
  S^{ij}(\mathbf{x}_\perp)
  =
  \frac{1}{4P^+}
  \int \frac{\mathrm{d}^2\boldsymbol{\Delta}_\perp}{(2\pi)^2}
  \Big(
  \boldsymbol{\Delta}_\perp^i \boldsymbol{\Delta}_\perp^j
  -
  \boldsymbol{\Delta}_\perp^2 \delta^{ij}
  \Big)
  D(t)
  e^{-i\boldsymbol{\Delta}_\perp\cdot\mathbf{x}_\perp}
  e^{-\frac{\sigma^2}{2} \boldsymbol{\Delta}_\perp^2}
  \,.
\end{align}
The pure stress tensor, notably, vanishes in the infinite momentum limit
(i.e., as $P^+\rightarrow\infty$),
but in any physically realistic frame (in which $P^+$ is finite),
the pure stress tensor is finite.
Because of the factor $\frac{1}{P^+}$, however,
the pure stress tensor is not invariant under longitudinal boosts,
and it is additionally not independent of state preparation.
Accordingly, any ``forces'' encoded in the pure stress tensor
are also frame- and state-dependent.
On the other hand, by multiplying by $P^+$,
one can obtain a frame- and state-independent quantity:
\begin{align}
  \tilde{S}^{ij}(\mathbf{x}_\perp)
  =
  P^+ S^{ij}(\mathbf{x}_\perp)
  =
  \frac{1}{4}
  \int \frac{\mathrm{d}^2\boldsymbol{\Delta}_\perp}{(2\pi)^2}
  \Big(
  \boldsymbol{\Delta}_\perp^i \boldsymbol{\Delta}_\perp^j
  -
  \boldsymbol{\Delta}_\perp^2 \delta^{ij}
  \Big)
  D(t)
  e^{-i\boldsymbol{\Delta}_\perp\cdot\mathbf{x}_\perp}
  e^{-\frac{\sigma^2}{2} \boldsymbol{\Delta}_\perp^2}
  .
\end{align}
The physical meaning of this expression is explored below through examples.

The pure stress tensor $S^{ij}(\mathbf{x}_\perp)$
is isotropic for spin-zero and spin-half particles,
as can be seen from examining Eq.~(\ref{eqn:Sij}).
This means that it can be parametrized by two independent functions
of $x_\perp^2$:
\begin{align}
  \label{eqn:Sij:ps}
  S^{ij}(\mathbf{x}_\perp)
  =
  \delta^{ij} p(x_\perp^2)
  +
  \left( \frac{x_\perp^ix_\perp^j}{x_\perp^2} - \frac{1}{2}\delta^{ij} \right)
  s(x_\perp^2)
  \,.
\end{align}
Here, $p(x_\perp^2)$ is the light front static pressure,
and the remaining components---parametrized by the function
$s(x_\perp^2)$---give the stress deviator tensor~\cite{Irgens:2008con},
whose components are shear stresses~\cite{Fetter:1980str,Irgens:2008con}.

A few conceptual remarks are in order regarding the physical meaning of
the pressure and shear forces in the light front stress tensor.
With $x^-$ integrated out, the description given by
$S^{ij}(\mathbf{x}_\perp)$ is inherently $(2+1)$-dimensional.
Surfaces in two-dimensional space are one-dimensional,
and accordingly the light front pressure has units of force/length
rather than force/area
(as can be confirmed through a unit analysis on Eq.~(\ref{eqn:Sij})).

\subsubsection{Properties of pressure and shear functions}

Although the Breit frame Fourier transform
does not actually give a spatial density,
a significant amount of work has been done using the Breit frame EMT.
It is thus instructive to construct analogies on the light front
to results that have already been obtained in literature using the
Breit frame.
Ref.~\cite{Polyakov:2018zvc} is especially helpful to compare to.

To begin, in analogy to Eq.~(23) of Ref.~\cite{Polyakov:2018zvc},
we find that the pressure and shear functions can be written:
\begin{subequations}
  \label{eqn:Dtilde}
  \begin{align}
    \widetilde{D}(x_\perp)
    &=
    \frac{1}{4P^+}
    \int \frac{\mathrm{d}^2\boldsymbol{\Delta}_\perp}{(2\pi)^2}
    D(t)
    e^{-i\boldsymbol{\Delta}_\perp\cdot\mathbf{x}_\perp}
    \\
    p(x_\perp)
    &=
    \frac{1}{2x_\perp}
    \frac{\mathrm{d}}{\mathrm{d}x_\perp}
    \widetilde{D}(x_\perp)
    +
    \frac{1}{2}
    \frac{\mathrm{d}^2}{\mathrm{d}x_\perp^2} \widetilde{D}(x_\perp)
    =
    \frac{1}{2x_\perp}
    \frac{\mathrm{d}}{\mathrm{d}x_\perp}
    \left[
      x_\perp
      \frac{\mathrm{d}}{\mathrm{d}x_\perp}
      \widetilde{D}(x_\perp)
      \right]
    \\
    s(x_\perp)
    &=
    \frac{1}{x_\perp}
    \frac{\mathrm{d}}{\mathrm{d}x_\perp} \widetilde{D}(x_\perp)
    -
    \frac{\mathrm{d}^2}{\mathrm{d}x_\perp^2} \widetilde{D}(x_\perp)
    =
    -
    x_\perp
    \frac{\mathrm{d}}{\mathrm{d}x_\perp}
    \left[
      \frac{1}{x_\perp}
      \frac{\mathrm{d}}{\mathrm{d}x_\perp}
      \widetilde{D}(x_\perp)
      \right]
    \,.
  \end{align}
\end{subequations}
The pressure in particular has another simple expression:
\begin{align}
  p(x_\perp)
  =
  \frac{1}{2}
  \delta_{ij} S^{ij}(\mathbf{x}_\perp)
  =
  -
  \frac{1}{8P^+}
  \int \frac{\mathrm{d}^2\boldsymbol{\Delta}_\perp}{(2\pi)^2}
  \boldsymbol{\Delta}_\perp^2
  D(t)
  e^{-i\boldsymbol{\Delta}_\perp\cdot\mathbf{x}_\perp}
  \,.
\end{align}
From this expression, it is clear that the 
von Laue stability condition~\cite{Laue:1911emt,Polyakov:2018zvc}:
\begin{align}
  \int \mathrm{d}^2\mathbf{x}_\perp \,
  p(\mathbf{x}_\perp)
  =
  0
  \,,
\end{align}
is automatically satisfied for any hadron.

Eqs.~(\ref{eqn:Dtilde}) show that $p(x_\perp)$ and $s(x_\perp)$ are
not independent functions.
In analogy to the discussion around Eq.~(30) of Ref.~\cite{Polyakov:2018zvc},
the non-independence of these functions can also be demonstrated through
the equilibrium condition $\nabla_i S^{ij}_{\mathrm{LF}}(\mathbf{x}_\perp) = 0$.
This condition entails:
\begin{align}
  p'(x_\perp) + \frac{1}{2} s'(x_\perp) + \frac{1}{x_\perp} s(x_\perp) = 0
  \,,
\end{align}
which is the 2D light front analogy to Ref.~\cite{Polyakov:2018zvc}'s Eq.~(30).
Here we use $p'(x_\perp)$ to denote differentiation with
respect to $x_\perp = |\mathbf{x}_\perp|$, i.e.,
\begin{align*}
  p'(x_\perp)
  =
  \frac{\mathrm{d}p(|\mathbf{x}_\perp|)}{\mathrm{d}|\mathbf{x}_\perp|}
  \,.
\end{align*}

Expressions for the light front pressure and shear densities---as well
as the normal and tangential force distributions---were previously
found in Sec.~IV of Ref.~\cite{Lorce:2018egm}.
Eqs.~(110,111) of \textsl{ibid.}\ in particular agree with 
Eqs.~(\ref{eqn:Dtilde}) of the current work,
up to the minor difference that $\frac{M}{P^+}$ was factored out of each
of these densities in Ref.~\cite{Lorce:2018egm}.
This work and Ref.~\cite{Lorce:2018egm} arrive at the same expressions
for different reasons, however.
The stress tensor contains a velocity flow piece with factors of
$\mathbf{P}_\perp$ that blows up for spatially-localized
($\mathbf{R}_\perp=0$) states.
In this work, we identified pressure and shear by looking at the
``pure stress tensor'', in which the velocity flow piece is removed.
In Ref.~\cite{Lorce:2018egm},
the EMT density was defined in phase space using a Wigner function formalism,
and the condition $\mathbf{P}_\perp=0$ was imposed throughout Sec.~IV.
This removes exactly the same terms from the stress tensor as we removed
from the pure stress tensor,
and since the remaining terms have no $\mathbf{P}_\perp$ dependence,
the results for the densities are the same.
We remark, however, that integration over all $\mathbf{P}_\perp$
is needed to obtain a physical density from a Wigner density,
and the condition $\mathbf{P}_\perp=0$ cannot be imposed
while integrating over $\mathbf{P}_\perp$.

\subsubsection{Stability conditions and mechanical radius}

As described above, the von Laue condition---that the integral of
pressure over all space is zero---is automatically satisfied for any hadron.
Another important stability condition is the
Polyakov-Schweitzer negativity condition~\cite{Polyakov:2018zvc},
which stipulates that the $x_\perp^2$-weighted moment of pressure should
be negative for any mechanically stable system.
We find the relevant moment to be:
\begin{align}
  \int \mathrm{d}^2\mathbf{x}_\perp \,
  \mathbf{x}_\perp^2
  p(\mathbf{x}_\perp)
  =
  \frac{1}{2P^+} D(0)
  \,,
\end{align}
meaning the Polyakov-Schweitzer condition can be written:
\begin{align}
  \label{eqn:negative}
  D(0) \leq 0
  \,.
\end{align}
This is identical to Eq.~(40) of Ref.~\cite{Polyakov:2018zvc}
(up to the possibility that $D(0)=0$, which is satisfied for instance
by free fermions).
Unlike the von Laue condition,
satisfying the negativity condition is not trivial,
and the requirement that $D(0) \leq 0$ can provide a useful sanity check
for model calculations of hadron structure.

A more general (non-trivial) stability condition can be derived,
following the discussion around Eq.~(39) of Ref.~\cite{Polyakov:2018zvc}.
The effective normal force density on a
1D surface---with units force/length---is given by
$S^{ij}(\mathbf{x}_\perp)\hat{n}_i$, where $\hat{n}$ is a unit vector
normal to the surface\footnote{
  It is in fact the pure stress tensor---the local value of the stress
  tensor as measured by a comoving observer---that corresponds to the
  normal force density.
  See Chapter 12 of Ref.~\cite{Fetter:1980str}.
}.
For a system centered at the origin, the requirement that normal forces
be directed strictly outwards---a stability condition protecting against
collapse---requires that
$S^{ij}(\mathbf{x}_\perp) \mathbf{x}_\perp^i \geq = 0$,
which imposes:
\begin{align}
  \label{eqn:positive}
  p(\mathbf{x}_\perp) + \frac{1}{2} s(\mathbf{x}_\perp) \geq 0
  \,,
\end{align}
which can be seen as a two-dimensional light front analogue of
Ref.~\cite{Polyakov:2018zvc}'s Eq.~(39),
and which was also previously found in Eq.~(128)
of Ref.~\cite{Lorce:2018egm}.
In terms of the function $\widetilde{D}(x_\perp)$ defined in
Eq.~(\ref{eqn:Dtilde}), this stability condition can be written:
\begin{align}
  \frac{1}{x_\perp} \widetilde{D}'(x_\perp) \geq 0
  \,,
\end{align}
which entails:
\begin{align}
  \label{eqn:negative:general}
  \int \frac{\mathrm{d}^2\boldsymbol{\Delta}_\perp}{(2\pi)^2}
  \left\{
    D(t)
    +
    t \frac{\mathrm{d}D(t)}{\mathrm{d}t}
    \right\}
  e^{-i\boldsymbol{\Delta}_\perp\cdot\mathbf{x}_\perp}
  \leq 0
  \,.
\end{align}
By integrating this condition over all space,
one obtains Eq.~(\ref{eqn:negative}) as a special case.
Eq.~(\ref{eqn:negative:general}) is however a stricter condition.

As noted in Ref.~\cite{Polyakov:2018zvc},
the normal force being positive (directed outwards)
means it can be used to define a mean squared radius,
which is called the mechanical radius in Ref.~\cite{Polyakov:2018zvc}.
Within the light front radius, we find the mechanical radius to be:
\begin{align}
  \label{eqn:radius:mechanical}
  \langle x_\perp^2 \rangle_{\mathrm{mech}}
  =
  \frac{
    \int \mathrm{d}^2 \mathbf{x}_\perp \,
    \mathbf{x}_\perp^2
    \Big[
      p(\mathbf{x}_\perp) + \frac{1}{2} s(\mathbf{x}_\perp)
      \Big]
  }{
    \int \mathrm{d}^2 \mathbf{x}_\perp \,
    \Big[
      p(\mathbf{x}_\perp) + \frac{1}{2} s(\mathbf{x}_\perp)
      \Big]
  }
  =
  \frac{
    4D(0)
  }{
    \int_{-\infty}^0 \mathrm{d}t \, D(t)
  }
  \,,
\end{align}
which is the light front analogue of
Ref.~\cite{Polyakov:2018zvc}'s Eq.~(41).


\section{Illustrations with a Dipole Parametrization}
\label{sec:models}

Several phenomenological parametrizations and
model calculations of the gravitational form factors
$A(t)$ and $D(t)$ for various mesons and baryons exist in the literature
(see Refs.~\cite{Masjuan:2012sk,Freese:2019bhb,Anikin:2019ufr,Neubelt:2019sou,Varma:2020crx} for instance).
Although it is not universal,
it is fairly common to use a multipole parametrization
to approximate the results of model calculations.
The generic multipole parametrizations take the form:
\begin{subequations}
  \label{eqn:multipole}
  \begin{align}
    A(t)
    &=
    \frac{1}{(1-t/\Lambda^2)^{n}}
    \\
    D(t)
    &=
    \frac{D(0)}{(1-t/\Lambda^2)^{n}}
    \,,
  \end{align}
\end{subequations}
where $n$ is the order of the multipole and $\Lambda$ is the multipole mass.
$n$ and $\Lambda$ can be different for $A(t)$ and $D(t)$,
but to reduce notational clutter we use the same notation for both.
We shall proceed to use these standard ansatzes
for the gravitational form factors
to illustrate the light front densities with the EMT.

\subsection{General formulas using the multipole form}

First, let's consider the light front momentum density,
which is given by Eq.~(\ref{eqn:density:p+}).
Using the multipole parametrization, we obtain:
\begin{align}
  \varepsilon(\mathbf{x}_\perp)
  =
  P^+
  \int \frac{\mathrm{d}^2\boldsymbol{\Delta}_\perp}{(2\pi)^2}
  \frac{
    e^{-i\boldsymbol{\Delta}_\perp\cdot\mathbf{x}_\perp}
  }{(1+\boldsymbol{\Delta}_\perp^2/\Lambda^2)^{n}}
  \,.
\end{align}
This integral can be approached by breaking $\boldsymbol{\Delta}_\perp$
into a component parallel to $\mathbf{x}_\perp$, and an orthogonal component,
and integrating out the latter.
This gives:
\begin{align}
  \varepsilon(\mathbf{x}_\perp)
  =
  P^+
  \frac{\Lambda^{2n}}{4\pi^2}
  \frac{\sqrt{\pi}\Gamma\left(n-\frac{1}{2}\right)}{\Gamma(n)}
  \int_{-\infty}^\infty \mathrm{d}k \,
  \frac{
    e^{-ikx_\perp}
  }{
    (\Lambda^2 + k^2)^{n-1/2}
  }
  \,.
\end{align}
Comparing to Basset's integral~\cite{NIST:DLMF} gives:
\begin{align}
  \label{eqn:multipole:e}
  \varepsilon(\mathbf{x}_\perp)
  =
  P^+ \Lambda^2
  \frac{(\Lambda x_\perp)^{n-1}}{2^{n}\pi\Gamma(n)}
  K_{n-1}(\Lambda x_\perp)
  \,,
\end{align}
where $K_n(x)$ is a modified Bessel function
of the second kind~\cite{NIST:DLMF}.

For the pure stress tensor, and the associated pressure and shear functions,
it is most straightforward to use $\widetilde{D}(x_\perp)$ as defined
in Eqs.~(\ref{eqn:Dtilde}).
In the multipole parametrization:
\begin{align}
  \widetilde{D}(x_\perp)
  =
  \frac{D(0)}{4P^+}
  \int \frac{\mathrm{d}^2\boldsymbol{\Delta}_\perp}{(2\pi)^2}
  \frac{
    e^{-i\boldsymbol{\Delta}_\perp\cdot\mathbf{x}_\perp}
  }{
    (1+\boldsymbol{\Delta}_\perp^2/\Lambda^2)^{n}
  }
  \,,
\end{align}
which can be evaluated as:
\begin{align}
  \widetilde{D}(x_\perp)
  =
  \frac{D(0)\Lambda^2}{4P^+}
  \frac{(\Lambda x_\perp)^{n-1}}{2^{n}\pi\Gamma(n)}
  K_{n-1}(\Lambda x_\perp)
  \,.
\end{align}
Using Eqs.~(\ref{eqn:Dtilde}) and recursion relations for the
modified Bessel functions~\cite{NIST:DLMF},
the pressure and shear functions are:
\begin{subequations}
  \label{eqn:multipole:ps}
  \begin{align}
    p(x_\perp)
    &=
    \frac{D(0)\Lambda^4}{4P^+}
    \frac{1}{2^{n+1}\pi \Gamma(n)}
    \Big\{
      (\Lambda x_\perp)^{n-1} K_{n-3}(\Lambda x_\perp)
      -
      2 (\Lambda x_\perp)^{n-2} K_{n-2}(\Lambda x_\perp)
      \Big\}
    \\
    s(x_\perp)
    &=
    -
    \frac{D(0)\Lambda^4}{4P^+}
    \frac{1}{2^{n}\pi \Gamma(n)}
    (\Lambda x_\perp)^{n-1} K_{n-3}(\Lambda x_\perp)
    \,.
  \end{align}
\end{subequations}
Here, in the cases $n=2$ (dipole form) and $n=1$ (monopole form),
the modified Bessel functions with negative index are equal
to the same function with a positive index.
Interestingly, although the pressure can flip signs
(and must, in order to satisfy the von Laue condition),
the shear function is strictly positive,
since $D(0)<0$ and since the modified Bessel functions of the second kind
are strictly positive.

The normal force is given by the LHS of Eq.~(\ref{eqn:positive}),
and in the dipole parametrization is:
\begin{align}
  F_n(x_\perp)
  =
  p(x_\perp) + \frac{1}{2} s(x_\perp)
  =
  -
  \frac{D(0)\Lambda^4}{4P^+}
  \frac{ (\Lambda x_\perp)^{n-2} }{2^{n}\pi \Gamma(n)}
  K_{n-2}(\Lambda x_\perp)
  \,.
\end{align}
Since $D(0) < 0$ and the modified Bessel function is positive,
$F_n(x_\perp) > 0$,
as expected of a system stable against collapse.

Using Eq.~(\ref{eqn:radius:p+}),
the light front momentum radius in the multipole parametrization is:
\begin{align}
  \langle x_\perp^2 \rangle_{\mathrm{mom}}
  =
  \frac{4n}{\Lambda^2}
  \,.
\end{align}
On the other hand, the mechanical radius for the multipole form can be found
using Eq.~(\ref{eqn:radius:mechanical}), giving:
\begin{align}
  \langle x_\perp^2 \rangle_{\mathrm{mech}}
  &=
  \frac{4(n-1)}{\Lambda^2}
  \,.
\end{align}

\subsection{Singular behavior for specific multipoles}

For $x\sim0$, the modified Bessel functions of the second kind
have the following limited form for positive $n$~\cite{NIST:DLMF}:
\begin{align}
  K_n(x) \sim \frac{1}{2} \Gamma(n) \left(\frac{1}{2}x\right)^{-n}
  \,.
\end{align}
By contrast, $K_0(x)$ diverges logarithmically as $x\rightarrow0$.

When the monopole form ($n=1$) is used for $A(t)$,
the light front momentum density
Eq.~(\ref{eqn:multipole:e}) is logarithmically singular at the origin.
This density remaining finite at the origin has been stipulated as
a stability condition, for instance in Ref.~\cite{Lorce:2018egm}.
However, a density that is singular at the origin seems to actually
occur in QCD,
in the case of the pion transverse charge density~\cite{Miller:2009qu}.
It is worth noting that while $\varepsilon(x_\perp)$ is singular in the
monopole case at $x_\perp=0$, the integral of $\varepsilon(x_\perp)$ over
any finite region of space---even one containing the origin---is strictly
finite.

Whether a transverse density can be singular at the origin remains
an open question.
However, lattice data suggest a monopole form for the
$A(t)$ form factor of the pion~\cite{Brommel:2007zz}.
Thus, it seems prudent to not too hastily discard the possibility
of transverse densities that are singular at the origin.

As noted in Ref.~\cite{Lorce:2018egm}, using the dipole form ($n=2$)
for the $D(t)$ form factor results in a shear function $s(x_\perp)$ that
is finite at $x_\perp=0$;
it also results in a pressure function $p(x_\perp)$ that is singular
at the origin.
These can be observed by looking at Eqs.~(\ref{eqn:multipole:ps}),
and recalling that $K_{-n}(x) = K_n(x)$ in these formulas.
In particular $s(0)\neq0$ would make the off-diagonal components of
$S^{ij}_{\mathrm{LF}}(0)$ diverge, as can be seen in Eq.~(\ref{eqn:Sij:ps}).

The requirement that all components of $S^{ij}_{\mathrm{LF}}(0)$ be finite
would thus rule out the dipole form for $D(t)$.
However, we do not see a fundamental reason why components of the
pure stress tensor must be finite at the origin.
Indeed, as is the case with $\varepsilon(x_\perp)$ with the monopole form,
the integral of
$S^{ij}_{\mathrm{LF}}(x_\perp)$
over any region of space---including one that contains the origin---is finite.
It seems prudent to not discard the possibility of a dipole form for $D(t)$,
though the singular densities this would entail should be kept in mind.

\subsection{Numerical illustrations}

We will now consider some specific values of the multipole form to illustrate
the properties of the light front densities.
As mentioned in Ref.~\cite{Masjuan:2012sk},
the meson dominance hypothesis suggests the presence of an $f_2$ pole in
the gravitational form factors.
We thus use $\Lambda=1.275$~GeV~\cite{Zyla:2020zbs}.
We consider $n=1,2,3,4$ (monopole, dipole, tripole, and quadrupole) forms.
Note that Eq.~(\ref{eqn:radius:mechanical}) excludes the monopole from
consideration for $D(t)$,
since it would result in an infinite integral for $F_n(x_\perp)$,
and thus a zero mechanical radius.
For the $D(t)$ form factor specifically,
we use $D(0)=-1$ for the purposes of illustration.
It is noteworthy that $D(0)\approx-1$ for the
pion~\cite{Novikov:1980fa,Voloshin:1980zf,Polyakov:2018zvc,Freese:2019bhb},
perhaps making these illustrations most illuminating for the pion specifically.

\begin{center}
  \begingroup
  \setlength{\tabcolsep}{7pt} 
  \renewcommand{\arraystretch}{1.5} 
  \begin{table}
    \caption{
      Values for the light front momentum and mechanical radii
      using the multipole ansatz in Eq.~(\ref{eqn:multipole}),
      with $\Lambda = 1.275$~GeV and $D(0)=-1$,
      for several orders of multipole.
    }
    \label{tab:values}
    \begin{tabular}{c c c c} \toprule
      $n$ &
      $\sqrt{\langle x_\perp^2 \rangle_{\mathrm{mom}}}$ &
      $\sqrt{\langle x_\perp^2 \rangle_{\mathrm{mech}}}$ &
      $F_n(0)$
      \\ \midrule
      $1$ &
      $0.310$~fm &
      $0$ &
      -- \\
      $2$ &
      $0.438$~fm &
      $0.219$~fm &
      $\infty$ \\
      $3$ &
      $0.536$~fm &
      $0.310$~fm &
      $0.338$~GeV/fm$^2$ \\
      $4$ &
      $0.619$~fm &
      $0.379$~fm &
      $0.113$~GeV/fm$^2$ \\
      \bottomrule
    \end{tabular}
  \end{table}
  \endgroup
\end{center}

First, we give numerical values for the light front and mechanical radii
in Tab.~\ref{tab:values}.
The monopole form gives a zero mechanical radius because the associated
``charge,'' in the denominator of Eq.~(\ref{eqn:radius:mechanical}),
is infinite for $n=1$.
For the dipole form, the radii are curiously equal;
however, it may turn out that $A(t)$ and $D(t)$ are multipoles of
different orders, or even that the effective multipole masses are different,
so one should not read too much into this coincidence.
Starting at $n=3$,

\begin{figure}
  \centering
  \includegraphics[width=0.32\textwidth]{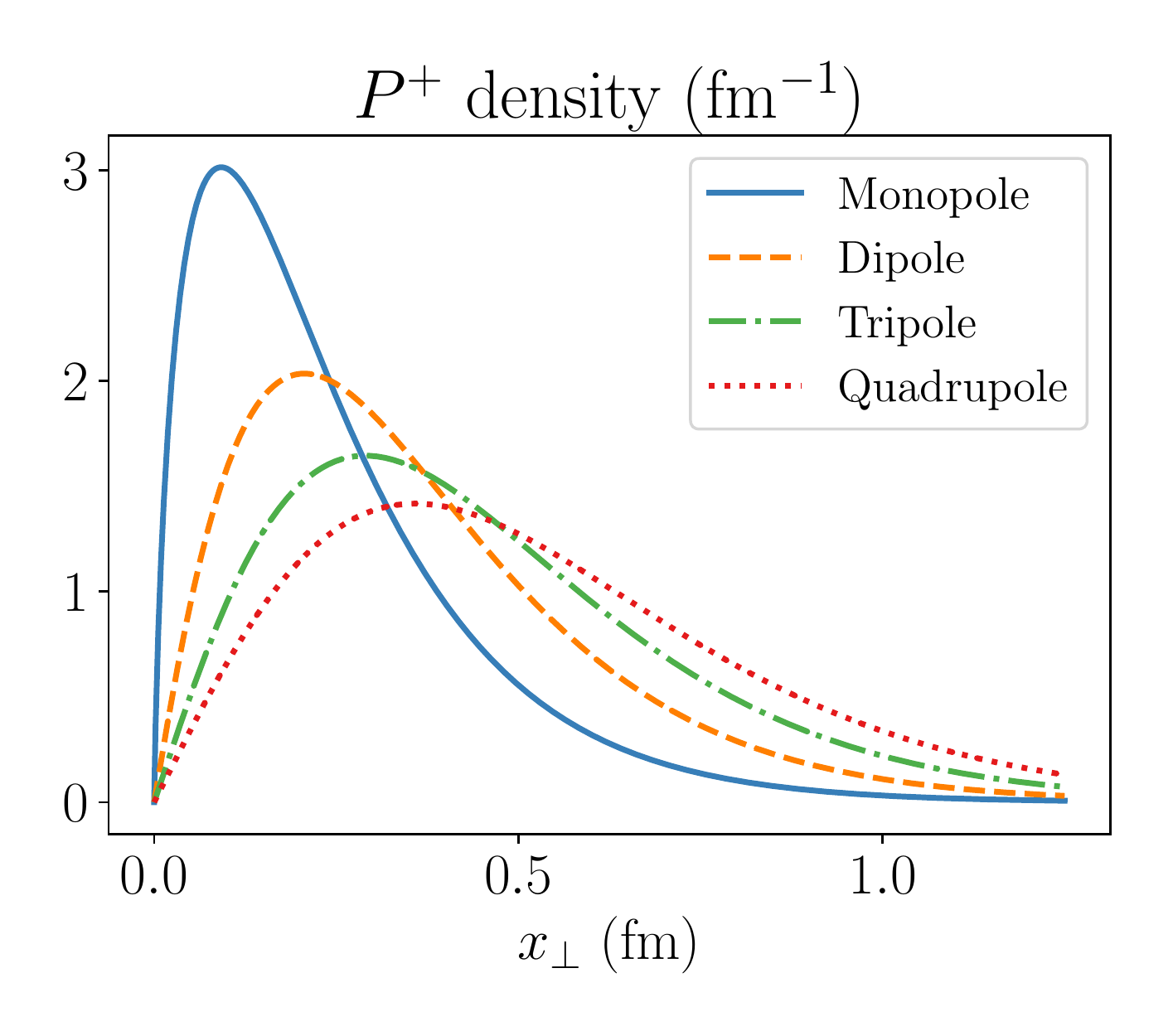} 
  \includegraphics[width=0.32\textwidth]{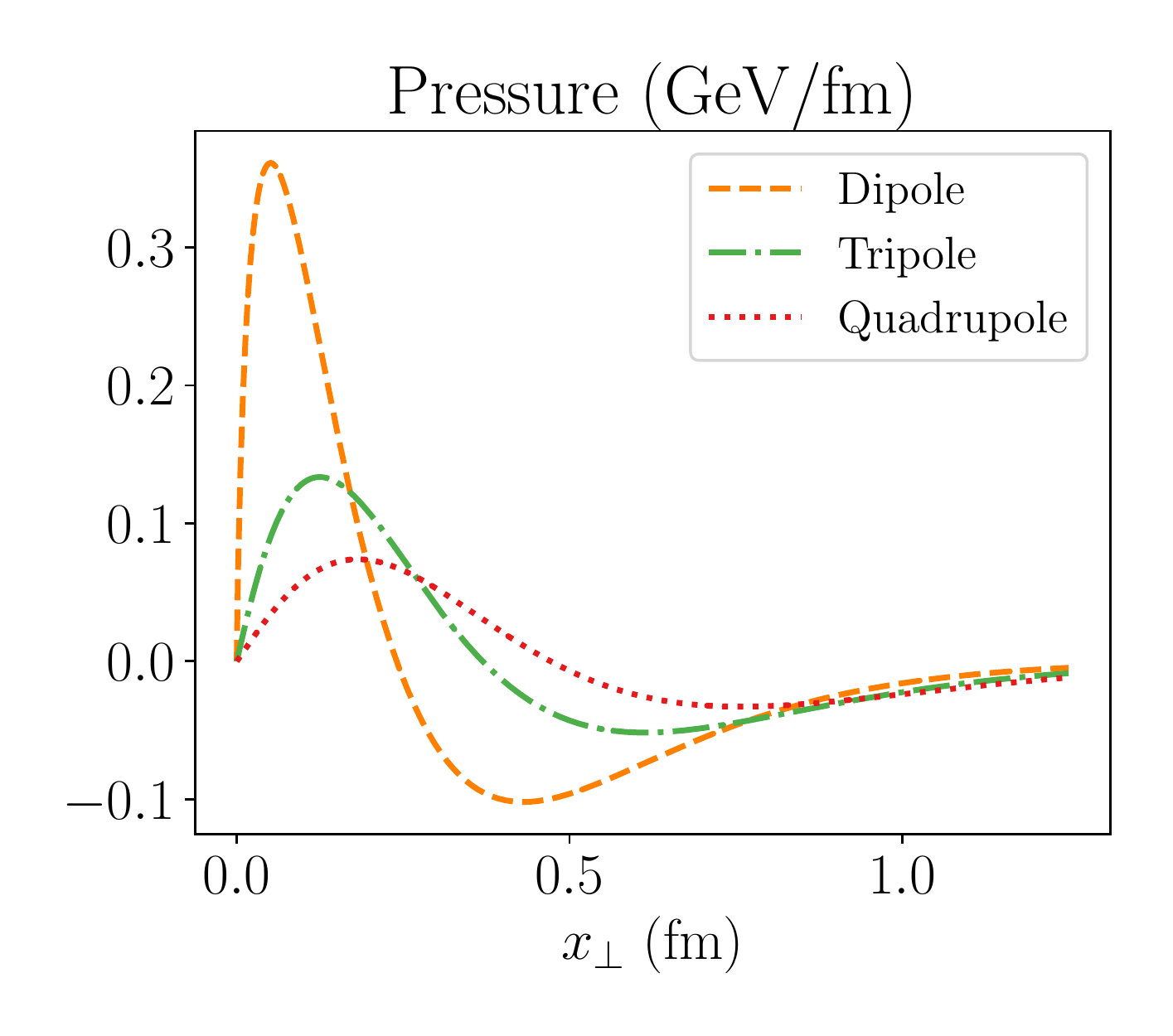} 
  \includegraphics[width=0.32\textwidth]{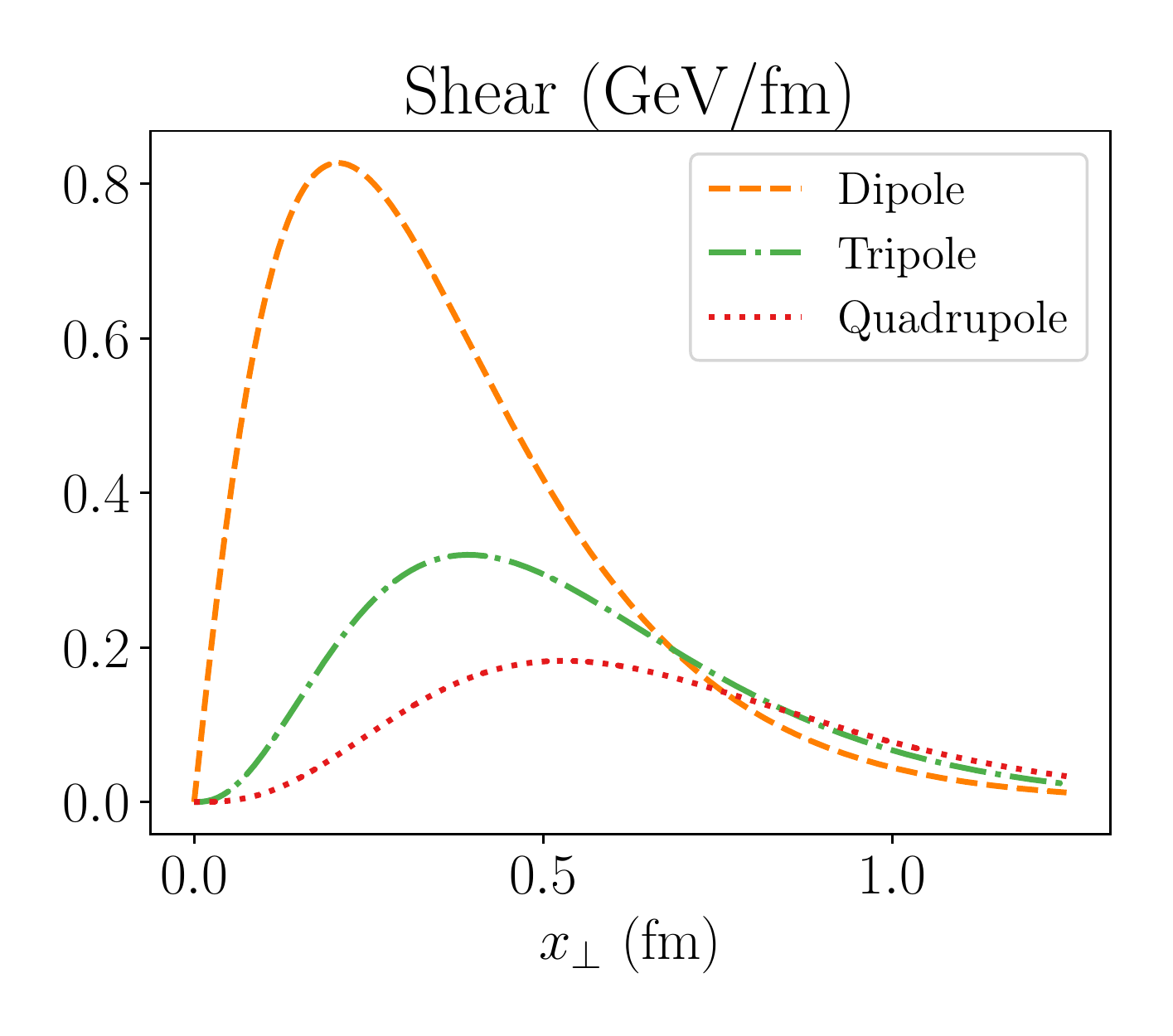} 
  \caption{
    Light front transverse densities associated with the EMT,
    for several multipole orders.
    Each density is weighted by $2\pi x_\perp$.
    The $P^+$ density is divided by $P^+$,
    while the pressure and shear are multiplied by $P^+$,
    to remove the $P^+$ dependence.
  }
  \label{fig:eps}
\end{figure}

The momentum, pressure, and shear densities are plotted in Fig.~\ref{fig:eps}
for several different orders of the multipole ansatz.
The $P^+$ density, with $P^+$ divided out,
integrates to $1$---a fact which is preserved by a conservation law.
The main effect of increasing the multipole order on this density is to make
the $P^+$ density more diffuse in the transverse plane,
a fact which is reflected in the increasing momentum radii
in Tab.~\ref{tab:values}.
By contrast, rather than merely becoming more diffuse,
the pressure and shear drop in magnitude when the multipole order is increased.

The magnitude of forces within the hadron decreasing with multipole order
is also reflected in Tab.~\ref{tab:values},
where the normal force at the center of the hadron decreases with
with increasing multipole order.
Thus, knowing only $D(0)$ is insufficient to know the magnitude of forces
(such as pressure) within a hadron;
it is necessary to know the full functional form of $D(t)$.

\begin{figure}
  \centering
  \includegraphics[width=0.5\textwidth]{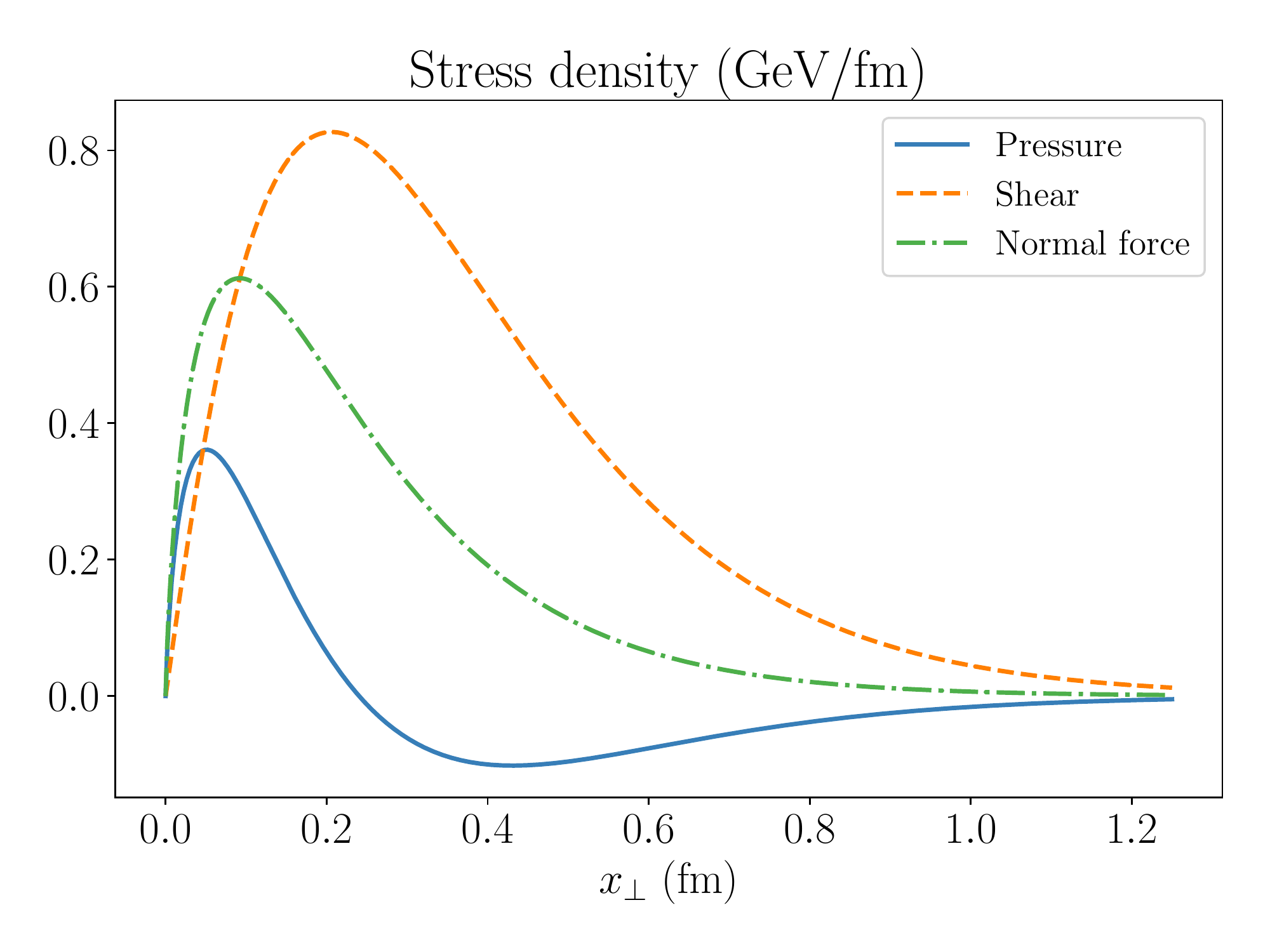} 
  \caption{
    Light front transverse densities associated with the
    pure stress tensor using the dipole ($n=2$) ansatz.
    Each density is weighted by $2\pi x_\perp$,
    and multiplied by $P^+$ to remove the $P^+$ dependence.
  }
  \label{fig:stress}
\end{figure}

In Fig.~\ref{fig:stress}, we plot the pressure, shear, and normal force
for the dipole ($n=2$) ansatz specifically.
In contrast to liquid drop models~\cite{Polyakov:2018zvc},
the shear function is not confined to a narrow region of space,
and is in fact quite broad, exceeding the pressure and the net
normal force in magnitude.
This finding was also observed for the multipole ansatz in
Ref.~\cite{Lorce:2018egm}.
On the one hand, the broadness of the shear function suggests that the hadron
cannot be seen as having a sharp boundary.
On the other hand, the large magnitude of the shear function suggests
the hadron is extremely viscous and cannot be interpreted as
an ideal fluid.
Of course, these qualitative conclusions only hold if the dipole ansatz
for $D(t)$ is in fact accurate.


\section{Conclusions and Outlook}
\label{sec:conclusion}

In this work, we derived the relativistically correct expression
for a hadron's transverse rest frame (``pure'') stress tensor.
The main result is given in Eq.~(\ref{eqn:Sij}),
which tells us how the D-term $D(t)$ encodes the spatial distribution of forces
in the transverse plane at fixed light front time.
Since the subgroup that keeps slices of fixed $x^+$ is Galilean,
the pure stress tensor $S^{ij}(\mathbf{x}_\perp)$ has the same formal properties
as the classical, non-relativistic stress tensor,
meaning that the classical laws of mechanics
(such as Cauchy's laws~\cite{Irgens:2008con})
can be applied.

Along the way, we obtained general expressions for spatial densities
within both the front form and instant form formalisms.
We found that in the instant form formalism,
one does not obtain a Breit frame Fourier transform.
Moreover, we reproduced the result of Ref.~\cite{Jaffe:2020ebz}
that the spatial extent of the hadron's wave function cannot be removed from
the density in the instant form formalism.

It is worth stressing that the results we have obtained were derived,
with a fundamental field-theoretic quantity---the expectation value
of a local operator with a physically realistic state---as
the starting point.
The spatial densities we have obtained were not postulated nor
defined by fiat,
and thus their interpretation is in clear and direct
correspondence with the actual spatial densities of hadrons.

In this work, we applied the formalism to spin-zero and spin-half hadrons,
and did not explore the spatial distribution of angular momentum.
Future work remains to be done on the distribution of angular momentum and
spin,
as well as the application of this formalism to spin-one particles,
which can have quadrupole moments~\cite{Cosyn:2019aio},
suggesting the stress tensor will not be isotropic.

\begin{acknowledgments}
  The authors would like to thank Julia Yu.\ Panteleeva
  for finding and correcting a factor 2 mistake in the mechanical radius
  results.
  This work was supported by the U.S.\ Department of Energy
  Office of Science, Office of Nuclear Physics under Award Number
  DE-FG02-97ER-41014.
\end{acknowledgments}


\bibliography{main.bib}

\end{document}